\documentclass[fleqn,10pt]{wlscirep}
\usepackage[utf8]{inputenc}
\usepackage[T1]{fontenc}

\title{Beneath the Surface: Revealing Deep-Tissue Blood Flow in Human Subjects with Massively Parallelized Diffuse Correlation Spectroscopy}

\author[1*]{Lucas Kreiss}
\author[1]{Melissa Wu}
\author[2]{Michael Wayne}
\author[1]{Shiqi Xu}
\author[3]{Paul McKee}
\author[3]{Derrick Dwamena}
\author[1]{Kanghyun Kim}
\author[1,4]{Kyung Chul Lee}
\author[5]{Wenhui Liu}
\author[2]{Arin Ulku}
\author[6]{Mark Harfouche}
\author[1]{Xi Yang}
\author[1]{Clare Cook}
\author[4]{Seung Ah Lee}
\author[7]{Erin Buckley}
\author[2]{Claudio Bruschini}
\author[2]{Edoardo Charbon}  
\author[3]{Scott Huettel}
\author[1,6]{Roarke Horstmeyer}

\affil[1]{Department of Biomedical Engineering, Duke University, Durham, NC 27708, USA}
\affil[2]{Advanced Quantum Architecture Laboratory, École polytechnique fédérale de Lausanne (EPFL), Neuchatel, NE 2000, Switzerland}
\affil[3]{Department of Psychology and Neuroscience, Duke University, Durham, NC, USA, 27708}
\affil[4]{School of Electrical \& Electronic Engineering, Yonsei University, Seoul, 03722, Republic of Korea}
\affil[5]{Department of Automation, Tsinghua University, Beijing, China}
\affil[6]{Ramona Optics, Inc., Durham, NC 27708, USA}
\affil[7]{Georgia Institute of Technology and Emory University, Wallace H. Coulter Department of Biomedical Engineering, Atlanta, Georgia, USA}

\affil[*]{\href{mailto:lucas.kreiss@duke.edu}{lucas.kreiss@duke.edu}}

\begin{abstract}
Diffuse Correlation Spectroscopy (DCS) allows label-free, non-invasive investigation of microvascular dynamics deep within tissue, like cerebral blood flow (CBF). However, the signal-to-noise ratio (SNR) in most conventional DCS techniques limits their use to pediatric applications, where the tissue between scalp and brain is substantially thinner as in adults. Here, we present the first \textit{in vivo} application of the most recent innovation in single photon detection technology to measure CBF in human adults at around 2~cm depth. We used an array with hundreds of thousands single photon avalanche diodes (SPAD) to boost the SNR by averaging all independent pixel measurements. The new system shows high blood flow sensitivity at an up to 2.6$\times$ increased depth, while maintaining a similar overall measurement noise as the previous state-of-the-art in parallelized DCS (4~cm source-detector-separation vs. 1.5~cm). Data from a large cohort of human adults provide strong experimental evidence for functional CBF activity during a cognitive memory task and the high temporal resolution allowed analysis of pulse markers under different conditions.
\end{abstract}
\begin{document}

\flushbottom
\maketitle
%
%
\thispagestyle{empty}

\begin{figure}[ht!]
\includegraphics[width=\textwidth]{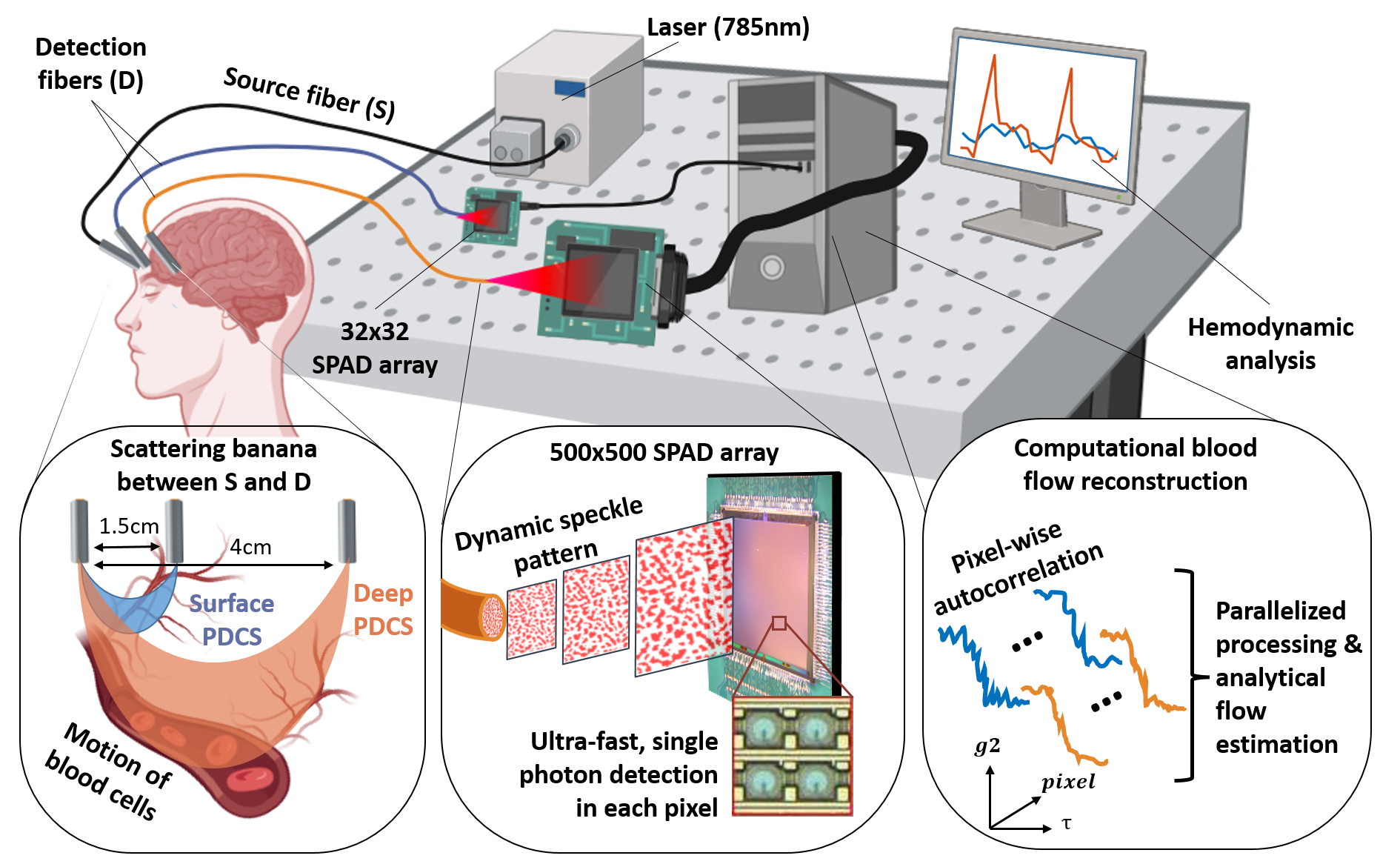}
\caption{Schematic of basic principle and experimental setup of parallelized diffuse correlation spectroscopy (PDCS). A coherent laser source is connected to an optical fiber which delivers light to the measurement location (e.g., the forehead for measurement of cerebral blood flow). Photons that were emitted by the source fiber (S) are then scattered diffusely within the tissue, and a small fraction finally reaches the detection fiber (D). A `banana' shaped probability distribution emerges over a great number of these individual optical paths between S and D. Photons that reached D are then transmitted to an array of single-photon avalanche diodes (SPAD). In our system, one setup (blue) is used at a short source detector separation (SDS, i.e., distance between S and D) to predominately measure photons that were scattered in superficial tissue layers. A second setup (orange) collects light at a larger SDS, which includes more photons that have traveled trough deeper tissue layers. Based on the random motion or diffusion of particles within the sample, a dynamically changing speckle pattern emerges and is measured by the high-speed SPAD array. Each pixel in the array samples a single speckle and thereby performs an independent DCS measurement. These individual intensity time traces are then used to compute autocorrelation curves, where the slope (i.e., decorrelation rate) can be linked to the speed of particle motion within the sample. By averaging all autocorrelations across thousands of pixels in the array, the signal-to-noise ratio can be boosted substantially. In biological tissues, the motion of blood cells can be measured using this technique and analytical models can be used to deduce hemodynamic metrics, like the blood flow index (BFI).}
\label{Fig_1}
\end{figure}

\section*{Introduction}
Diffuse correlation spectroscopy (DCS) is based on the scattering of coherent light at moving particles. In biological tissues, the dynamics of red blood cells reflect a primary example of highly scattering particles that are moving. DCS is still effective to measure blood flow dynamics at great tissue depths (>~1~cm) and does not rely on any artificial contrast agents, which makes it a valuable, label-free tool for quantifying blood flow in a noninvasive fashion. DCS is especially appealing for applications in the field of neuroscience, with the aim to measure cerebral blood flow (CBF) beneath the barrier of the scalp and skull. In a typical DCS setup, coherent laser light illuminates the sample at a defined location, while scattered photons are collected at a known distance to this illumination spot. It is often straight-forward to use optical fibers for illumination and detection and place them at a well-defined distance to each other (see Fig.~\ref{Fig_1}). Photons that are emitted by the source (S) are scattered randomly multiple times within the tissue, a certain fraction of which finally reach the detector (D). The summation of all photon paths between S and D leads to a stochastic scattering profile, with a typical `banana-shape', where a larger source detector separation (SDS) leads to greater measurement depth (see Fig.~\ref{Fig_1} and Fig.\ref{Fig_2}~b~\&~c). The effective depth of this scattering profile has been approximated to be 1/3 or 1/2 of the SDS~\cite{durduran2010diffuse}. Most commonly, DCS has been used at SDS~=~2~-~2.5~cm)~\cite{lee2019noninvasive,shoemaker2023using,lin2019_DCS_FPGA_Phatom,baker2017_DCS_forehead,choe2014_DCS_breast_cancer}, relating to a measurement depth of $\sim$ 1~-~1.25~cm. Compared to functional near-infrared spectroscopy (fNIRS), which is based on absorption instead of scattering, DCS has been reported to be more sensitive to CBF~\cite{selb2014sensitivity,baker2014modified} and the DCS metric of blood flow index (BFI) has been shown to have a better contrast-to-noise ratio during brain activation than absorption in NIRS~\cite{zhou2021functional}. However, the average thickness of scalp and skull in human adults is at the same order of magnitude (i.e., $1.48 \pm 0.28$~cm~\cite{wu2022complete}), as the typical DCS measurement depth ($\sim$ 1~-~1.25~cm). Thus, most conventional DCS techniques are usually applied to measure CBF in infants and children rather than in adults, where the extracerebral barrier is much thinner.

Nevertheless, traditional DCS systems already contributed significantly to understanding blood flow dynamics in various clinical contexts, including Subarachnoid hemorrhage (SAH)~\cite{sathialingam2023microvascular}, sickle cell disease (SCD)~\cite{lee2019noninvasive,Cowdrick2023_pediatric_sickle_cell_disease}, and delayed cerebral ischemia (DCI)~\cite{sathialingam2023microvascular}. DCS showed significant differences in the BFI of patients that suffered strokes~\cite{mesquita2011_DCS_clinical_validations_review} or in brain areas associated with speech of people who stutter~\cite{tellis2011_DCS_speech_tasks} to name a few recent examples from the ever growing body of research. Despite these successful applications, conventional DCS systems are still constrained to a certain measurement depth. 

Many different strategies have been used to address this limitation, to increase the signal-to-noise ratio (SNR) and the sensitivity of DCS, and ultimately to increase its measurement depth while maintaining high temporal sampling rate. Interferometric setups, for instance, are sensitive to faster speckle variations, which is generally advantageous for DCS~\cite{zhou2018highly,xu2020_interferometric_DCS,zhou2021functional,zhao2023interferometric}. Interferometric diffusing wave spectroscopy (fiDWS) has been reported at 4~cm SDS and 10~Hz sampling rate or even at 5~cm SDS and 0.1~Hz sampling rate, albeit from a relatively low number of five subjects~\cite{zhou2021functional}.

A different, and rather straight-forward method to increase SNR is the simultaneous measurement of multiple, independent speckles. In 2007, Dietsche et al. used a fiber bundle, an array of detectors and a multichannel autocorrelator to show DCS at 2.9~cm SDS from a single subject~\cite{dietsche2007fiber}. A similar concept has been used for `multi-speckle DCS'~\cite{murali2020_MDCS_Forehead} or `multiplexed DCS' with an 5$\times$5 array~\cite{johansson2019multipixel}. This concept of parallelization has also made its way into interferometric setups, like the `parallel, interferometric diffusing wave spectroscopy'~\cite{zhou2018highly}. These systems measure $\sim$ 20~\cite{zhou2018highly} - 25~\cite{johansson2019multipixel} speckles in parallel and by averaging all $M$ independent autocorrelation curves, the SNR can consequently be increased by a factor of $\sqrt{M}$. A combined strategy of using a longer wavelength of 1,064nm, an interferometric setup and multi-speckle detection by four superconducting nanowire single-photon detectors (SNSPD) was deployed by Robinson et al. to measure cerebral blood flow at 100~Hz sampling rate and a SDS of 3.5~cm from five subjects~\cite{robinson2023portable}. 

The recent development of CMOS SPAD arrays with many hundreds and thousands of individual SPAD pixels is fueling this ongoing trend. For instance, Sie et al.~\cite{sie2020high} as well as Liu et al.~\cite{liu2021_parallel_DCS_forehead_and_phantoms} used a 32$\times$32 SPAD array and averaged all 1,024 pixels to boost the SNR 32-$\times$ compared to a single pixel measurement, terming this technique multispeckle DCS~\cite{sie2020high} or parallelized DCS (PDCS)~\cite{liu2021_parallel_DCS_forehead_and_phantoms}. In this work, we will use the later term PDCS from here on. A similar PDCS setup with a 32$\times$32 SPAD array was then used to classify spatiotemporal-decorrelating patterns deep beneath turbid media~\cite{xu2022transient} and to exploit such patterns to reconstruct images of flow patterns~\cite{xu2022_DCS_imaging}. However, it should be noted, that the effective SNR-gain in SPAD arrays over individual SPADs is usually slightly below the theoretical factor of $\sqrt{M_{pixel}}$, since each pixel in the array has a lower photon detection efficiency (PDE) as conventional, single SPADs (e.g., PDE $\sim$ 15\% at 785~nm in arrays against up to 70~\% in single SPADs)~\cite{james2023diffuse}. Most recently, a large-format SPAD camera with 500$\times$500 has been demonstrated~\cite{wayne2022500}. Compared to the 1,024 pixels in our previous PDCS systems~\cite{liu2021_parallel_DCS_forehead_and_phantoms,xu2022_DCS_imaging,xu2022transient}, this new SPAD array has a total of 250,000 pixels and therefore offers a massively increased potential for parallelized DCS measurements. Optical phantom experiments already showed that this new SPAD array can be used for PDCS experiments, following the same $\sqrt{M_{pixel}}$ gain in SNR, boosting it by a factor of up to 473 compared to a single SPAD pixel in the array~\cite{wayne2023massively}. Despite this very promising technological advance, SPAD cameras of this massive size have not yet been utilized for \textit{in vivo} experiments to study actual blood flow in human subjects.

While the promise of these new DCS technologies for SNR gain are evident, most of these proof-of-concept studies were only carried out on optical phantoms~\cite{wayne2023massively} or on a very limited number of human subjects (n), e.g., n=1~\cite{dietsche2007fiber}, n=7~\cite{choe2014_DCS_breast_cancer}, n=3~\cite{murali2020_MDCS_Forehead}, n=4~\cite{zhou2021functional}, n=6~\cite{liu2021_parallel_DCS_forehead_and_phantoms}, or n=5~\cite{robinson2023portable}, which prevents or limits a statistical significance analysis of blood flow metrics and SNR behavior over a larger number of human subjects and under real experimental conditions.

In this publication, we demonstrate the first use of the latest 500$\times$500 SPAD camera for massively parallelized DCS in a large cohort of human subjects and under real experimental conditions of neuro-cognitive studies. We show that this new PDCS system offers the ability to optically measure fast blood flow dynamics at great depth of approximately 2~cm (SDS~=~4~cm), while still enabling fast sampling of pulsatile blood flow at 8~-~10~Hz. We carried out a systematic study with over 15 human subjects, investigating global suppression as well as localized activation of blood flow. A dual-detection system with two different PDCS systems - one at short SDS and one at large SDS - further allowed a direct, in-built reference between superficial and deeper tissue layers.

\section*{Results}
In this work, we explored massively parallelized DCS in two different \textit{in vivo} experiments. In the first experiment, we initiated a global decrease of blood flow in the forearm via a pressurized arm cuff and measured PDCS blood flow with our dual-detection system with two different measurement depths. In the second experiment, we placed the same system on the forehead to compare  the localized blood flow activity in deeper brain tissue with the more superficial measurement from scalp tissue as a reference. For more details on study design, optical setup, data processing and statistical analysis, please refer to the methods section as well as to our supplementary material. 

\subsection*{Sensitivity to deep and superficial blood flow during global suppression in the forearm}

\begin{figure}[ht!]
\centering
\includegraphics[width=0.8\textwidth]{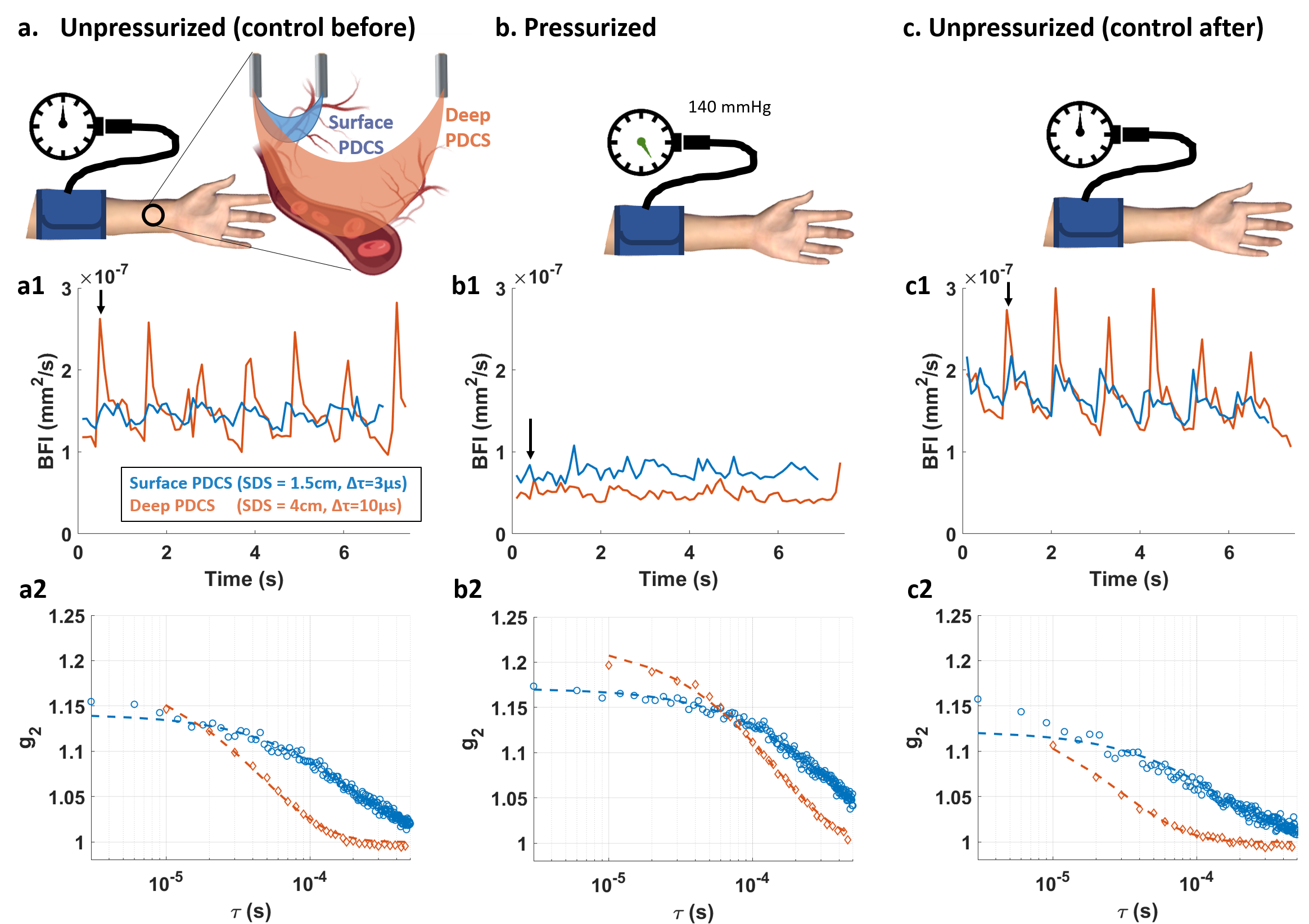}
\caption{PDCS measurement at the arm of an exemplary subject before (a), during (b), and after (c) blood flow suppression via a pressurized cuff. (a1,b1,c1) show the blood flow index (BFI). The BFI was sampled at 10~Hz and each individual value was derived from fitting an analytical model to the averaged autocorrelation curves ($g_2 (\tau)$). (a2,b2,c2) show one exemplary $g_2 (\tau)$ that was used to obtained the BFI marked by an arrow in (a1,b1,c1). Orange color indicates the deep PDCS measurement (SDS = 4~cm, $\Delta \tau = $10~\textmu s, averaged across $\Delta T = 0.1$~s and 125,000 pixels). Blue color indicates the superficial PDCS measurement (SDS = 1.5~cm, $\Delta \tau = $3~\textmu s and averaged across $\Delta T = 0.1$~s and 1,024 pixels).}
\label{Fig_3}
\end{figure}

In the first experiment, we initiated a global decrease of blood flow in the forearm via a pressurized arm cuff and measured PDCS blood flow with our dual-detection
system. For both PDCS configurations (SDS~=~1.5~cm and SDS~=~4~cm), the derived BFI values (Fig.~\ref{Fig_3}~a1, c1) clearly show the characteristic diastolic and systolic peaks that are typically present in arteries~\cite{clark2018BloodFlow}, which confirms blood flow sensitivity for superficial and deep tissue measurements. As expected, the $g_2$ curves decay much faster in the PDCS measurements at larger source-detector separation compared to shorter source-detector separation under the same conditions, since the longer photon path leads to more scattering events.  Upon suppression of blood flow via a pressurized arm cuff (Fig.~\ref{Fig_3}~b), the $g_2$ curves show a later decay and the derived BFI time traces show a significantly decreased BFI value overall and no pulsatile behavior, indicating the suppression of blood cells motion. 

\begin{figure}
\centering
\includegraphics[width=0.75\textwidth]{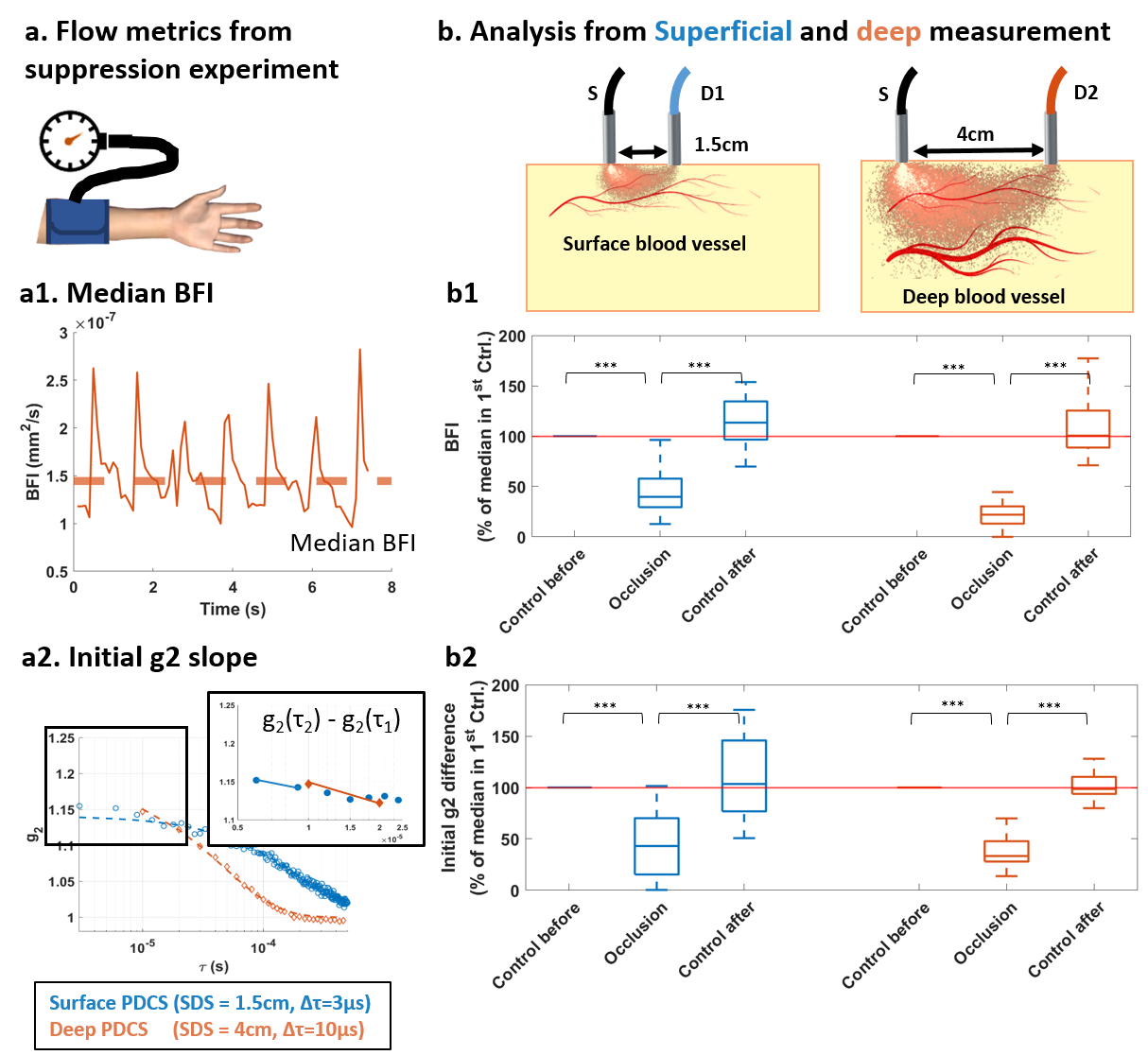}
\caption{Group data analysis of PDCS measurement at the arm (blood flow suppression via pressurized cuff). (a) The median BFI value and the median of the initial $g_2$ slope were derived for each individual trial. (b) Results of all trials were compared between the three conditions (control before, test, control after) and for both detection schemes (superficial PDCS at SDS=1.5~cm in blue and deep PDCS at SDS=4cm in organge). Both metrics (BFI in b1 and initial $g_2$ slope in b2) show a statistically significant decrease during blood flow suppression ($*p<0.05$ for all cases). All boxplots show the 25th and 75th percentile as boxes. Significance levels are indicated as $^{*}p<0.05, ^{**}p<0.01, ^{***}p<0.001$. Data include 33 independent trials from 11 subjects. }
\label{Fig_5}
\end{figure}


Fig.~\ref{Fig_5} shows the grouped PDCS results from all trials in the blood flow suppression experiment. These results confirm that the effect in an individual trial, described above in (Fig.~\ref{Fig_3}, also holds over a larger group of different subjects. For the superficial PDCS measurement, the suppression of blood flow in the forearm via a pressurized cuff resulted in an average decrease of 50~\% in normalized BFI (Fig.~\ref{Fig_5}~b1), when compared to the first control. Upon release of the pressure (`Control after'), the normalized BFI reached an average of around 120~\%, indicating that the normalized BFI exceeded the initial level during reactive hyperemia (i.e., when the blood rushed back). For the deep tissue PDCS measurement, this change is more pronounced, reaching an average decrease of 75~\% during suppression when compared to the control before or an average difference of 83~\% when compared to the control after. The results were statistically significant in all conditions (p < 0.001). 

In addition to the more rigorous and well-established BFI (Fig.~\ref{Fig_5}~a1,~b1), we tested the initial $g_2$ slope as blood flow metric (Fig.~\ref{Fig_5}~a2). This metric follows the same general trend in the group statistics as the BFI described above - albeit at a different magnitude (see Fig.~\ref{Fig_5}~b2). For the superficial tissue measurement in the forearm, this metric shows a highly significant decrease of 46~\% and 63~\% when compared to the control before or after, respectively. For the deep tissue PDCS measurement in the same experiment, the average decrease of this metric is 62~\% and 63~\%. Again, these results were statistically significant in all conditions and at both measurement depths (p < 0.001). 

\newpage

\subsection*{Sensitivity to localized, cerebral blood flow activity}

\begin{figure}[ht]
\centering
\includegraphics[width=0.8\textwidth]{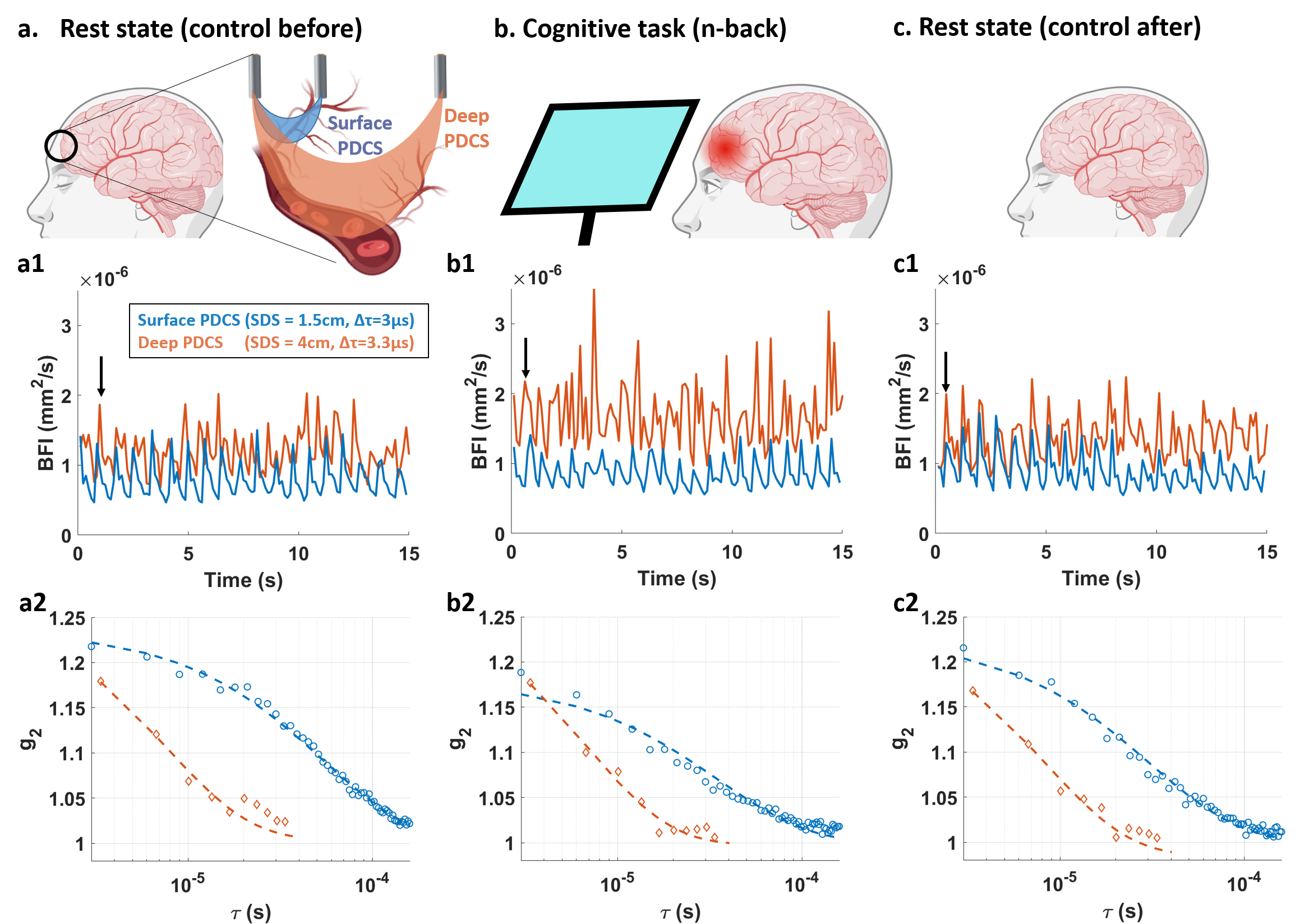}
\caption{PDCS measurement at the prefrontal cortex of an exemplary subject before (a), during (b), and after (c) the cognitive task of an n-back test. (a1,b1,c1) show the blood flow index (BFI). The BFI was sampled at 8~Hz and each individual value was derived from fitting an analytical model to the averaged autocorrelation curves ($g_2 (\tau)$). (a2,b2,c2) show one exemplary $g_2 (\tau)$ that was used to obtained the BFI marked by an arrow in (a1,b1,c1). Orange color indicates the deep PDCS measurement (SDS = 4~cm, $\Delta \tau = $3.3~\textmu s and averaged across $\Delta T = 0.125$~s and 64,000 pixels). Blue color indicates the superficial PDCS measurement (SDS = 1.5~cm, $\Delta \tau = $3~\textmu s and averaged across $\Delta T = 0.125$~s and 1,024 pixels).}
\label{Fig_4}
\end{figure}

In the second experiment, we placed the system on the forehead to measure blood flow activity during a cognitive task that is known to induce activity in the prefrontal cortex. Similar to the previously reported results from the forearm, the $g_2$ curves from the forehead generally decay much faster at larger SDS (see Fig.~\ref{Fig_4}~a3, b3, c3), the derived BFI values (Fig.~\ref{Fig_4}~a1, b1, c1) show pulsatile blood flow at both configurations (SDS~=~1.5~cm and SDS~=~4~cm) and the peak pulse BFI is higher for the deep PDCS measurement compared to the BFI at superficial PDCS recordings. In contrast to the results from the previous suppression experiment, the $g_2$ curves at larger SDS experience more noise, especially towards longer delay values, as seen in Fig.~\ref{Fig_4}~a3, b3, c3.  In this experiment, the derived BFI values (Fig.~\ref{Fig_4}~a1, b1, c1) are generally higher for longer source-detector separation, which might indicate cerebral sensitivity, since it has been reported that BFI can be up to 6 to 10 times higher in the brain as in the surrounding tissue~\cite{selb2014sensitivity}. During the active cognitive task (Fig.~\ref{Fig_4}~b) the BFI at large source-detector separation (SDS~=~4~cm) clearly shows an overall increase compared to the control measurement, while the BFI that was simultaneously measured at SDS~=~1.5~cm shows little to no effect during this task. As shown in Fig.~\ref{Fig_2}~c, this trend is expected if brain sensitivity was reached, as a much larger portion of scattered photon will penetrate scalp and skull at an SDS of 4~cm, as compared to an SDS of only 1.5~cm. 

\begin{figure}
\centering
\includegraphics[width=0.95\textwidth]{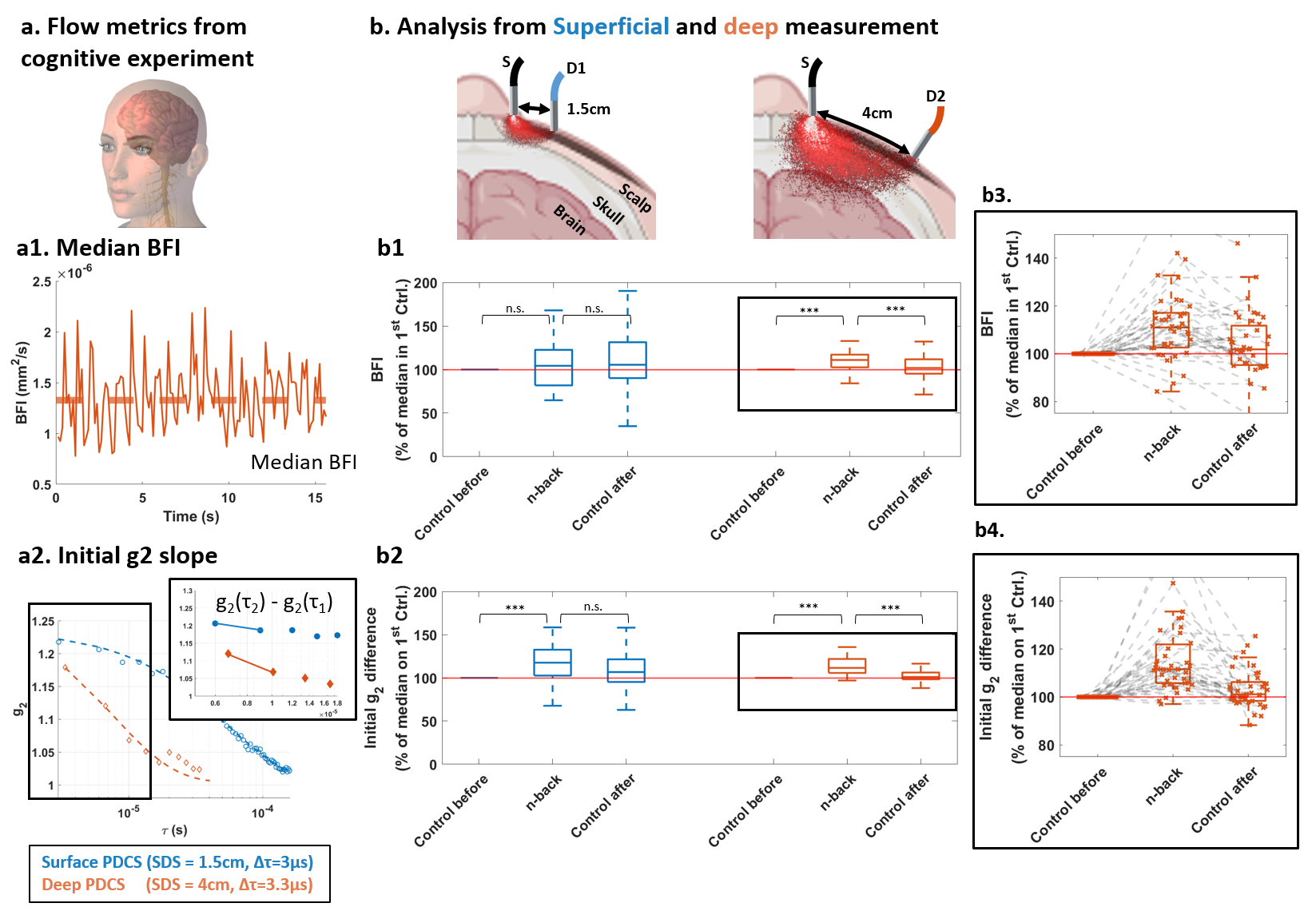}
\caption{Group data analysis of PDCS measurement at the prefrontal cortex (blood flow activation during cognitive task). (a) The median BFI value and the median of the initial $g_2$ slope were derived for each individual trial. (b) Results of all trials were compared between the three conditions (control before, n-back, control after) and for both detection schemes (superficial PDCS at SDS=1.5~cm in blue and deep PDCS at SDS=4cm in organge). Only the deep PDCS configuration shows a statistically significant increase during the cognitive task in both metrics, indicating cerebral sensitivity. (b3~\&~b4) show re-scaled versions of the group differences from the deep tissue layer (at SDS~=~4~cm). All boxplots show the 25th and 75th percentile as boxes. The gray dotted lines in (b3~\&~b4) show that most trials experienced an increase during the task and again an decrease afterwards, although few individual trials did not yield this result. Significance levels are indicated as $^{*}p<0.05, ^{**}p<0.01, ^{***}p<0.001$. Data include 39 independent trials from 15 subjects.}
\label{Fig_6}
\end{figure}

Fig.~\ref{Fig_6} shows the respective group statistics for the cognitive activation experiment. These results from 15 subjects show the same trend as the BFI in an individual trial, described above in Fig.~\ref{Fig_4}. For the superficial PDCS measurement, the change in normalized BFI did not reach statistical significance in any condition (p > 0.05), which indicates little to no change in BFI at the more superficial tissue layers. In contrast to that, the deeper PDCS measurement at SD~=~4~cm reaches statistical significance for the grouped BFI results (p < 0.001),  with an average increase of 12~\% and 8~\%, when compared to the control before or after the cognitive task, respectively (see Fig.~\ref{Fig_6}~b1,~b3). Thus, our experiment provides experimental evidence that PDCS at 4~cm SDS can reach cerebral sensitivity in human adults.

In the cognitive experiment, the initial $g_2$ slope (Fig.~\ref{Fig_6}~a2) shows a similar trend as the BFI. Data from the superficial PDCS configuration shows statistical significance of this metric, when comparing the n-back test to the control before with an average increase of 20~\% (p < 0.001). However, no significant change was detected when comparing the test to the control after (p > 0.05). In the deep PDCS configuration, the data show a statistically significant increase in both conditions (p < 0.001). The average change is 15~\% and 12~\%, when compared to the control before or after, respectively (Fig.~\ref{Fig_6}~b2,~b4). 

Although this trend is apparent and statistically significant, the magnitude of the change in the initial $g_2$ slope of >60~\% for the arm or >15~\% in the PFC should \textit{not} be mistaken for an equivalently large difference in flow. The magnitude of this change is not linearly proportional to actual blood flow changes, since it might be affected by several other parameters, like changes in absorption (e.g., due to increased hemoglobin volume), fluctuations $\beta$ and others. In order to investigate if this difference was indeed driven by an increased hemoglobin volume, we measured the average speckle intensity, which would indicate absorption changes, similar to a conventional NIRS measurement. As seen in the supplementary material, this intensity metric did not indicate a distinct difference of statistical significance, when comparing cognitive activation with the control cases. 

\subsection*{Pulsatility analysis}

\begin{figure}[ht]
\centering
\includegraphics[width=1\textwidth]{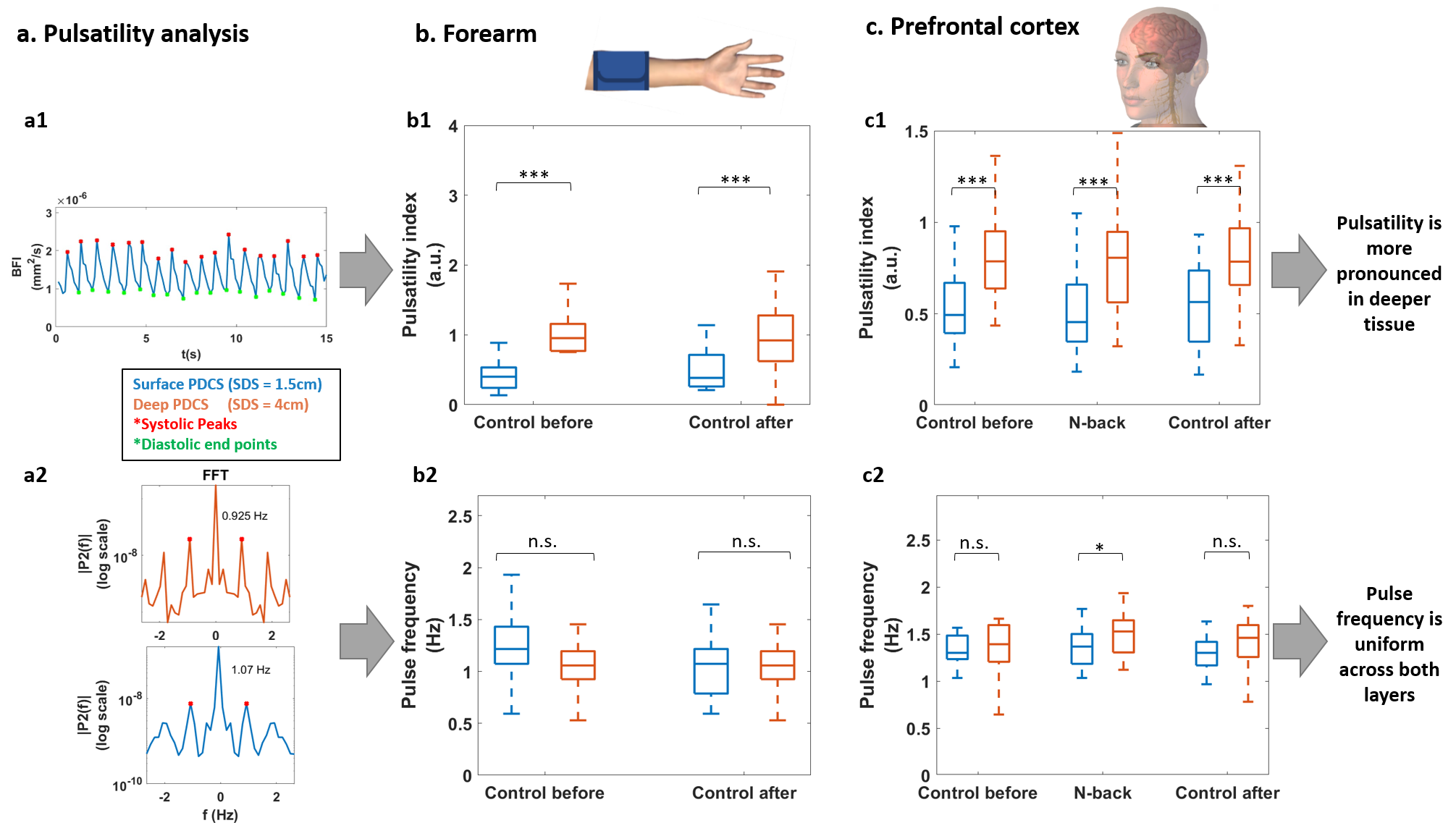}
\caption{Pulse analysis. (a) Our algorithm detects systolic peaks, diastolic endpoints, diastolic notches and diastolic peaks in data from both SDS (a1), as well as the pulse rate  (a2). Based on these metrics, the pulsatility index (b1, c1) was calculated and analyzed statistically across all measurements for the suppression experiment in the forearm (b), as well as for the cognitive experiment (c). Data during suppression of blood flow in the arm is not shown, since there was no puslatile flow in this case. Significance levels are indicated as $^{*}p<0.05, ^{**}p<0.01, ^{***}p<0.001$. Data include 33 independent trials from 11 subjects in (b) and 39 trails from 15 subjects in (c).}
\label{Fig_7}
\end{figure}

A merely qualitative inspection of the BFI time traces (Fig.~\ref{Fig_3}~\&~\ref{Fig_4}, as well as Fig.~\ref{Fig_7}~a) suggests that the BFI from the superficial PDCS measurement shows lower BFI peak values than the deep PDCS measurement. This qualitative observation was tested quantitatively, by detecting the  pulse markers of systolic peaks, diastolic endpoints, diastolic notches and diastolic peaks in the pulsatile BFI traces, as shown in Fig.~\ref{Fig_7}~a1. Based on these pulse markers, we calculated the pulsatility index (PI), as described in the methods section (additional metrics of resistance index and notch index are shown in the supplementary material). Additionally, a fast Fourier transform (FFT) was carried out to obtain the overall pulse frequency of each trial (Fig.~\ref{Fig_7}~a2). Fig.~\ref{Fig_7}~b~\&~c show the statistical analysis of these parameters across all trials, comparing the superficial measurement (SDS~=~1.5~cm, blue) to the deeper measurement (SDS~=~4~cm, orange).

In the case of the global flow suppression in the arm, shown in Fig.~\ref{Fig_7}~b, the PI (Fig.~\ref{Fig_7}~b1) showed clearly significant increases in the measurement at greater SDS as compared to the reference at lower SDS. The PI was increased by an average of 0.44~-~0.49 in both control measurements (Cohen's D = 1.1~-~1.2, $p<10^{-4}$). The pulse rate however (Fig.~\ref{Fig_7}~b2), showed no notable differences between the two simultaneous detection schemes and all data within $25^{th} - 75^{th}$ percentile were within a range of 0.8~-~1.4~Hz or 48~-~84 bpm (Cohen's D = -0.3~-~0.1, $p>0.2$), which relates well to the expected physiological range for resting heart rate in healthy adults of about 1~-~1.3~Hz (60~-~80~bpm)~\cite{clark2018BloodFlow}. The test condition during blood flow suppression was excluded, since there was no pulsatile flow during suppression.

The results for PI in the PFC experiment (Fig.~\ref{Fig_7}~c1) generally follow a similar trend of significantly increased pulsatility at greater SDS. The PI was increased by 0.28 (Cohen's D of 1.1~-~1.2, $p<10^{-4}$) for all three conditions. The measured pulse rate (Fig.~\ref{Fig_7}~c3) showed that all data within $25^{th} - 75^{th}$ percentile were within a range of 1.1~-~1.4~Hz or 66~-~84 bpm (Cohen's D<0.5, p>0.01) and showed no significant difference for the two control cases. During the cognitive task, our data slightly reached statistical significance with an average increase of 0.15~Hz or 9~bpm higher between lower SDS and larger SDS (Cohens D~=~0.53, $p<0.03$). 

\subsection*{Noise analysis}

\begin{figure}[ht]
\centering
\includegraphics[width=0.8\textwidth]{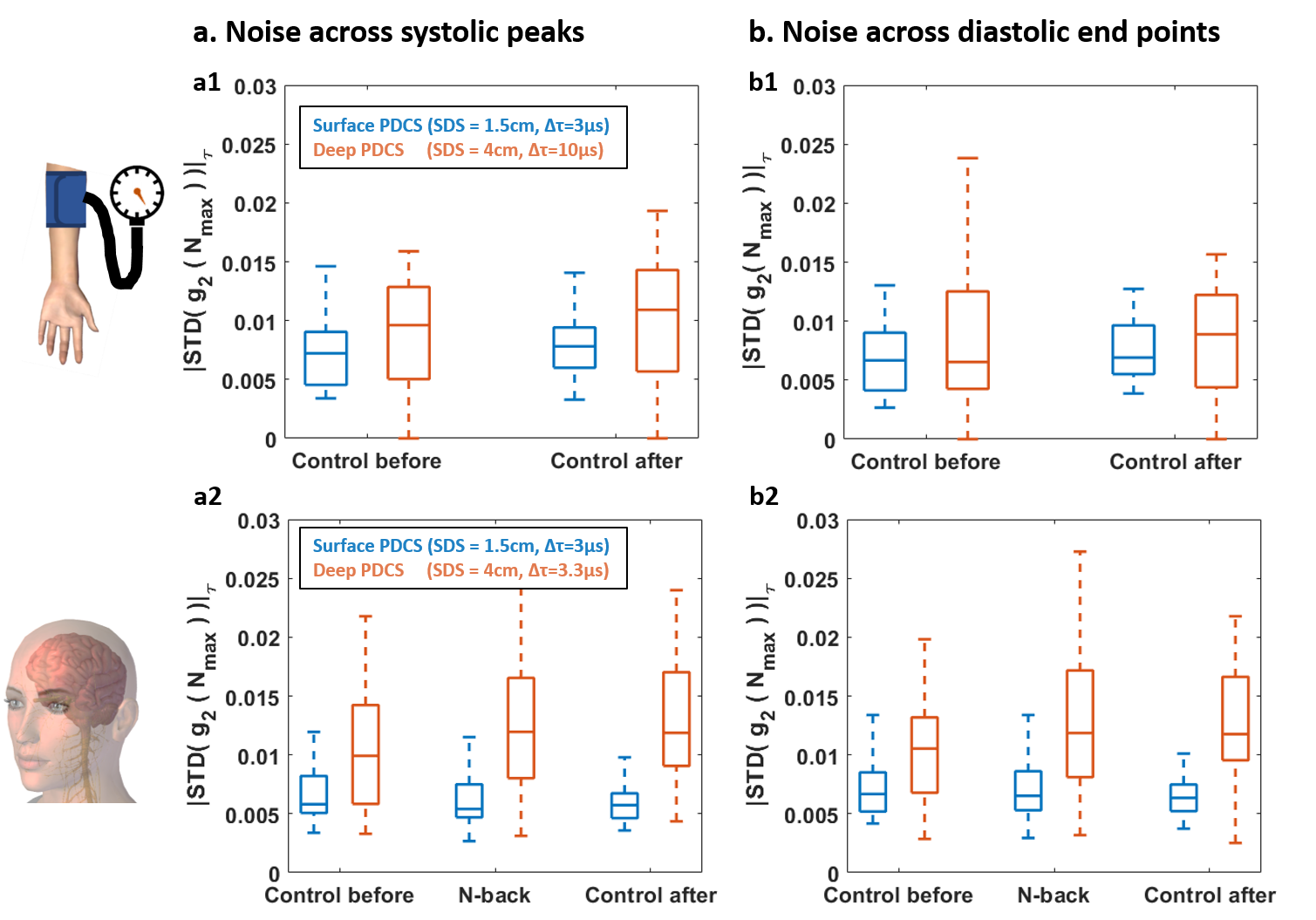}
\caption{The new 500$\times$500 array can enhance the measurement depths in PDCS by achieving a similar noise level at 4~cm SDS, as the previous PDCS with a 32$\times$32 array at only 1.5~cm SDS. Experimental noise was calculated as the standard deviation (STD) of the $g_2$ curves across all subsequent systolic peaks (a) or diastolic endpoints (b). Results from the global blood flow suppression in the forearm (a1~\&~b1) show that the noise from the 32$\times$32 array at 1.5~cm SDS and 3~\textmu s exposure is on the same level as that of the larger detector with 125,000 pixels at 4~cm SDS and 10~\textmu s exposure. In the experiment on local activation in the PFC (a2~\&~b2) the noise reduction of the larger detector was deliberately scarified by recorded only 64,000 pixels to achieve a shorter exposure time of 3.3~\textmu s and thus a better flow sensitivity at the cost of increased noise (smaller $\Delta \tau$). Data include 33 independent trials from 11 subjects for forearm and 39 trails from 15 subjects for the cognitive experiment.}
\label{Fig_8}
\end{figure}

The detection of pulse markers shown above was also leveraged to facilitate a noise analysis in our \textit{in vivo} PDCS measurements. The noise in DCS can be calculated as the standard deviation (STD) of the $g_2$ curves across subsequent measurements under the same experimental conditions. In past experiments, this has been carried out in optical phantom experiments under "static" conditions (i.e., only considering a constant diffusion coefficient in the medium)~\cite{sie2020_DCS_SPADarray,liu2021_parallel_DCS_forehead_and_phantoms,wayne2023massively}. Based on the previously discussed detection of systolic peaks and diastolic endpoints (Fig.~\ref{Fig_7}~a), we calculated the noise in our data as the STD across subsequent systolic peaks (maximal flow, Fig.~\ref{Fig_8}~a) or diastolic endpoints (minimal flow, Fig.~\ref{Fig_8}~b) in each trial. This noise metric can be interpreted as the holistic combination of all possible sources of noise, including technical \textit{and} biological variations in the measurements, across subsequent blood flow peaks. In Fig.~\ref{Fig_8}, the distribution of this metric is then plotted across all trials. 

Results from the global blood flow suppression in the forearm (Fig.~\ref{Fig_8}~a1~\&~b1) show that the noise of the detector with 1,024 pixels at 1.5~cm SDS and 3~\textmu s exposure is on the same level as that of the larger detector with 125,000 pixels at 4~cm SDS and 10~\textmu s exposure. In the experiment on local activation in the PFC (Fig.~\ref{Fig_8}~a2~\&~b2) the noise reduction of the larger detector was deliberately scarified by recorded fewer pixels to achieve a shorter exposure time and thus a better sampling of the autocorrelation curves (smaller $\Delta \tau$). In this case, the median noise of the larger detector at 4~cm SDS is generally about two or three times as high as that of the  smaller detector at 1.5~cm SDS. Please refer to our supplementary material for more details, as well as a pixel-level noise analysis.

\section*{Discussion}

This study leveraged the latest developments in cutting-edge SPAD camera technology for parallelized Diffuse Correlation Spectroscopy (PDCS) for the first \textit{in vivo} experiments with a robust sample size. While previous the state-of-the art in \textit{in vivo} PDCS applications used much smaller array with only 1,024 parallelized detections, we showcase the use of 64,000 to 125,000 parallelized detection for a much more substantial boost in SNR. This allowed us to record deep-tissue, pulsatile blood flow with an SDS of 4~cm and a sampling rates of 8~-~10~Hz, which is on par with the current front-runners in DCS technology~\cite{zhou2021functional,robinson2023portable}, and applying it in a large study to cerebral blood flow activity in human adults. To the best of our knowledge, this is the first study of this kind.

In contrast to most previous studies on new DCS prototypes, we present a large study cohort of over a dozen adult subjects under real experimental conditions, which allowed to perform robust statistical analysis of all of our findings. In summary, our system is sensitive to global suppression of blood flow, as well as to a localized blood activation in the brain, which was validated by in-built reference measurement at lower SDS. Furthermore, we were able to provide strong experimental evidence for an increase pulsatility and resistance in deeper PDCS measurements (greater SDS), and we could quantify the overall noise in these measurements, including technical and biological variations. 

\paragraph{Trade-off between flow-sensitivity and SNR} The fidelity of massively parallelized DCS is governed by the trade-off between temporal resolution of the autocorrelation function $g_2(\tau)$ and the signal-to-noise ratio (SNR) in photon detection. In PDCS, the exposure time of each SPAD pixel defines the temporal sampling of these $g_2$ curves ($\Delta \tau$), which is essential for deep-tissue measurements, as the average number of scattering events increases towards many thousands, which causes the temporal speckle fluctuations to increase to the MHz regime. The shutter of the SwissSPAD3 array defines the total exposure time by the number of rows that are read out. More rows allow more parallelized measurements and thus higher SNR, but result in longer exposure times, which increases $\Delta \tau$ and thereby inhibits sensitivity to faster and deeper flow. 

Our results from the arm experiment showed that it is possible to acquire PDCS measurements with 125,000 pixels (250 rows from one sensor half for 250x500 pixels) from the new swissSPAD3 array at 4~cm SDS at a similar noise level as the conventional 1,024 pixel array at 1.5~cm SDS (note: the diameter of the active sensor area is similar for both arrays with 6.9~\textmu m for PF-32 and 6~\textmu m for swissSPAD3, temporal averaging window was identical at 0.1~s in both detection schemes). However, this came at the cost of reduced temporal resolution of the $g_2$ curves for the larger array of 10~\textmu s compared to the 3~\textmu s of the smaller PF-32. Nevertheless, both detectors still adequately sampled the autocorrelation decay the resulting BFI drop during flow suppression in the forearm.

The issue of reduced sensitivity to deeper and faster flow caused by longer integration time $\Delta \tau$ and thus a more sparse sampling of the autocorrelation, becomes much more critical for the measurement of cerebral blood flow in the prefrontal cortex (PFC), where the flow is naturally faster as in the sourounding tissue~\cite{selb2014sensitivity}. Therefore, we employed a new onboard autocorrelation calculation of only 64,000 pixels (64 rows from each of the two sensor halves for 128x500 pixels) to reduce the $\Delta \tau$. In order to slightly compensate for the respective reduction in SNR, we used a larger temporal averaging window $\Delta T$ of 0.128~s and fewer decay values per $g_2$ curve to average more independent autocorrelation curves. Although this procedure reduced the temporal sampling rate of the BFI trace, it still allowed detection and analysis of pulsatile blood flow, while maintaining similar exposure times for both arrays ($\Delta \tau = $3.37~\textmu s at 4~cm SDS vs. 3~\textmu s at 1.5~cm SDS).

These settings allowed us to measure clearly pulsatile blood flow in both experiments and to perform a more granular analysis of pulsatile metrics (Fig.~\ref{Fig_7}), as compared to the mere median blood flow metrics shown in Figures~\ref{Fig_5}~\&~\ref{Fig_6}

\paragraph{Dual detection in PDCS: "A bunch of bananas"}
The utilization of two different detectors at distinct SDS facilitated the comparison of blood flow changes at shorter SDS (superficial tissue layers) with those at greater SDS that also reach deeper tissue. Thereby, the detector at short SDS served as an in-build reference measurement under the exact same experimental conditions. As expected, both detectors show a clear drop in blood flow during global suppression in the forearm. However, only the deep PDCS configuration at greater SDS was sensitive to cerebral blood flow changes during the cognitive task, while the reference measurement at lower SDS showed no significant change. This is expected, if cerebral sensitivity was reached, since only the deeper scattering banana at greater SDS can reach brain tissue, wile the more shallow measurement at lower SDS would mostly detect extracerebral scattering in scalp and skull. 

Thereby, our dual detection scheme allowed to rule out other potential causes for increased blood flow, not assigned to cerebral activity, such as eye motion or motion artifacts, while also allowing a noise estimation for both configurations. While similar approaches of using multiple detection systems for different measurement depths have been already been used in DCS~\cite{shoemaker2023using,lee2019noninvasive,cowdrick2023agreement}, its application in PDCS with entire arrays of SPADs is new. 

Attempts to subtract the extracerebral contributions from scalp and skull in deep measurements are commonly used for fNIRS and some attempts have already been made to translate this approach to DCS~\cite{cowdrick2023agreement,Nakabayashi2023}. General linear models (GLM), for instance, have been used to regress out the relative changes in BFI at short SDS (i.e., scalp hemodynamics) from those at long SDS (i.e., hemodynamics in scalp and brain). However, the models assume a linear relation between superficial and deeper measurements, which is generally valid for fNIRS but not for DCS. While fNIRS is based on changes in optical absorption based on blood volume, DCS is fundamentally based on the motion of blood cells. This velocity of blood cells however, differs non-linearly across arteries, arterioles, capillaries, and veins, due to the differences in vessel diameter, wall thickness, and overall flow resistance~\cite{clark2018BloodFlow}. Therefore, it has been reported that such multi-layer models "should be taken with caution"~\cite{zhao2023comparison} and that homogeneous models perform better under general assumptions~\cite{zhao2023comparison}. Our new experimental evidence on pulsatility and resistance indices in this study, provide clear experimental evidence for higher resistance and higher pulsatility at greater depth. We interpret this to be plausible, since superficial measurements include a higher ratio of smaller capillaries (slower flow velocity, lower pulsatility), while the deeper PDCS measurement at SDS~=~4~cm reached more of the larger arteries in deeper tissue layers, that have higher pulsatility. Since the relation between the average blood cell velocity in capillaries and in larger vessels is not linear~\cite{clark2018BloodFlow}, the use of GLM for this PDCS was not feasible for our data.

Alternatively, more complex multi-layer diffusion correlation models have been explored for DCS, where each layer of tissue is modeled as a separate slap with distinct optical properties~\cite{cowdrick2023agreement}. However, these models are generally more mathematically complex (e.g., including a Bessel function) and include a greater number of input parameters (e.g., distinct optical properties in each slap), as the simpler semi-infinite model with homogeneous optical properties. Therefore, these models require more data points per autocorrelation function to reach the same level of certainty and to prevent overfitting. Since this number of delay values per autocorrelation function was limited by the exposure time of the array in our experimental data (e.g., 46 or 168 delay values in the arm experiment and 15 or 56 in the PFC respectively, see supplementary material), these models did not yield reliable results for BFI estimation.
In the future, more sophisticated data processing techniques might be able to leverage such dual-detection PDCS data to correct the large-SDS measurement by the extracerebral background and obtain an informed estimation of the signal that arises purely from the cerebral tissue at greater depth. We believe that these techniques could include additional information about the vasculature at the respective tissue depth, as well as mathematical models on how this vasculature shapes the flow dynamics. 

\paragraph{Limitations and challenges} The study acknowledges potential limitations and challenges in terms of the underlying biology, the technical setup, and the data processing. With respect to the technical setup, we recognize the limitations in SNR-vs-flow sensitivity trade-off that were already discussed above. Additionally, there is a natural variance in all biological parameters, like scalp and skull thickness, pulse rate or pulse-to-pulse variations. For instance, the average thickness of scalp and skull has been reported at approx. $1.48 \pm 0.28$~cm~\cite{wu2022complete}, while the depths of the scattering banana at SDS~=~4~cm is roughly 1.3~-~2~cm. Under these considerations, the number of photons that reach brain tissue would vary significantly between different subjects of different scalp and skull thickness and thus, our described methodology might yield varying cerebral sensitivity for each individual subject. This is supported by our results, where the median BFI did not indicate cerebral sensitivity for every individual trial, although the general group difference did show a notable effect (see zoomed inlet of Fig.~\ref{Fig_6}~b). 

Finally, some measurements were discarded based on a pre-defined exclusion criterion  according to the goodness of the BFI model fit to the data. A potential explanation for a reduced fit accuracy in these cases could be a miss-match between the assumed optical properties, that relied on reasonable estimations commonly used in literature (see supplementary material), and the actual properties of the respective subject. The absorption parameter $\mu_a$, in particular, has been described to vary largely between different skin color types (i.e., a 5-$\times$ difference in melanin concentration)~\cite{visscher2017skin} and separate values for each individual subject should be taken into account for future experiments.

\paragraph{Outlook} Already in 2020, a one Mpixel SPAD camera was reported, albeit at a relatively low frame rate of 24~kfps~\cite{morimoto2020megapixel}. This study used a SPAD camera with 0.25 Mpixel at 97~kfps (or fewer pixels at higher rate)~\cite{wayne2022500}. It is very likely that the ongoing development of SPAD camera technology will result in larger and faster detectors in the future, which would allow to sample the $g_2$ curves at higher temporal resolution (smaller $\Delta \tau$) for higher blood flow sensitivity, while also allowing more parallelized DCS measurements to increase the SNR even further. These advances would allow higher sampling rates (smaller averaging windows) of the blood flow metrics, higher temporal resolution of blood flow dynamics and ultimately even deeper DCS measurements. In addition to that, new data processing techniques, tailored for the sparse signal of binary photon detection events in SPAD detectors are an anticipated refinement for the future. This study provides experimental evidence in using massively parallelized SPAD cameras to measure cerebral blood flow \textit{in vivo} and therefore lays the groundwork towards advancing physiological research and clinical diagnostics in the future.

\section*{Methods}
The exposure time of a SPAD array is essential for DCS, as it defines the minimal sampling of the autocorrelation curve ($\Delta \tau$). Our massively parallelized swissSPAD3 is operated by 2 field programmable gate arrays (FPGAs) - one per sensor half and the number of readout rows in each row can be adjusted between 1 and 250. Readout of a single row corresponds to an exposure time of 40~ns~\cite{wayne2022500,wayne2023massively} and the full-frame exposure time is therefore defined as $t_{exposure}~=~n_{rows} \times 40$~ns. This configuration essentially results in a fundamental trade-off between the temporal resolution of the $g_2$ curves, on the one hand, and the number of independent pixel measurements and thus SNR increase, on the other hand. This trade-off is summarized in table~1 by Wayne et al.~\cite{wayne2023massively}. We explored this trade-off in two different experiments.

\subsection*{Study design}

\begin{figure}[ht]
\centering
\includegraphics[width=0.75\textwidth]{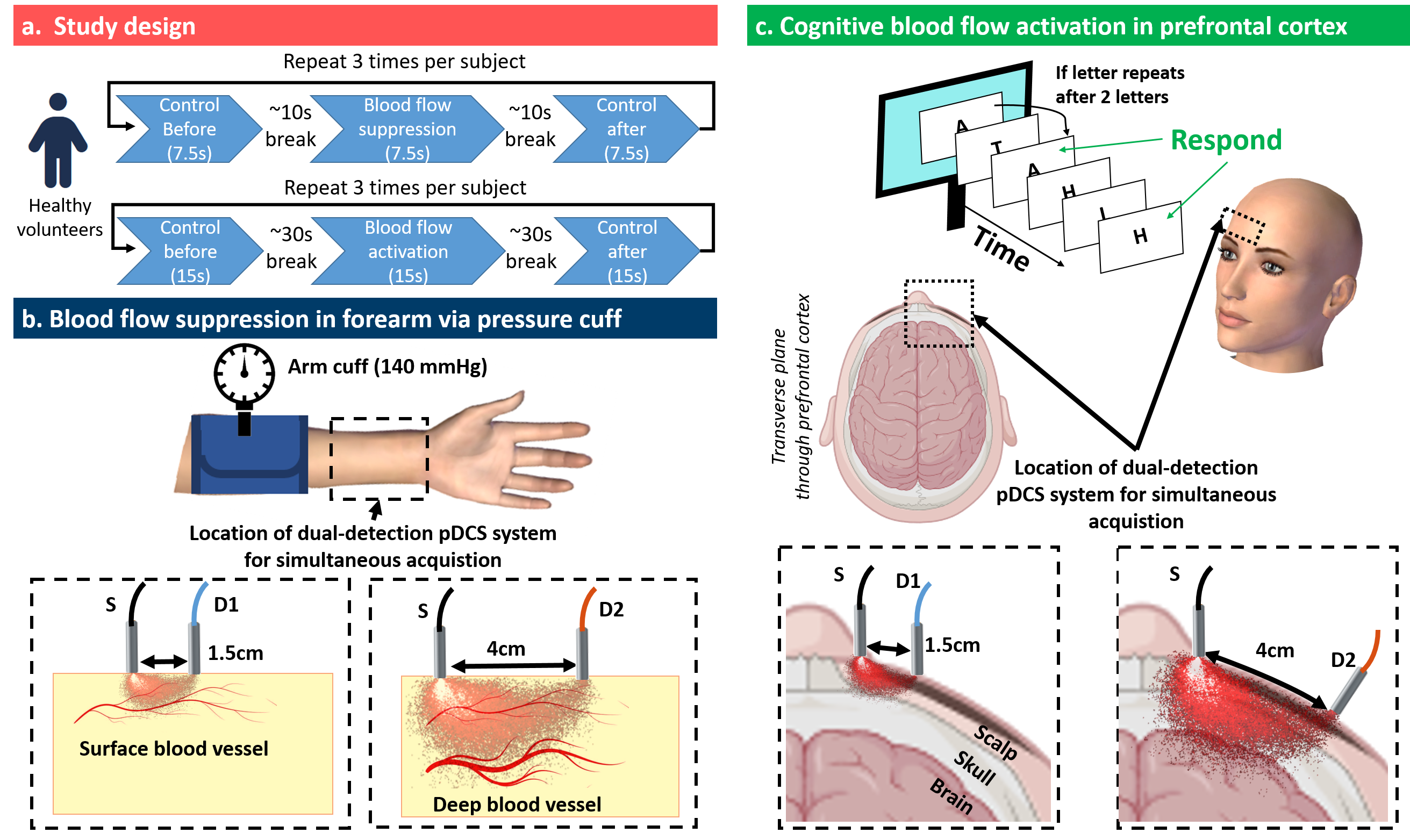}
\caption{Study design (a) and principle of dual-detection with PDCS (b) at the forearm and (c) the forehead.}
\label{Fig_2}
\end{figure}

Healthy adult subjects (n~=~24) were recruited for this study. To explore deep-tissue PDCS, data was collected from two different locations, under two different conditions: (i) general blood flow suppression in the entire forearm and (ii) localized activation of blood flow within brain tissue at the prefrontal cortex (PFC). At the forearm, the test condition was flow suppression via a blood pressure cuff and PDCS data was collected before, during, and afterwards. In the second experiment, subjects participated in a cognitive task (see below) and PDCS was recorded at the forehead, above the PFC before, during, and after this task. The same procedure was repeated three times for three independent measurements per subject, as displayed in Fig.~\ref{Fig_2}~a. We refer to these independent measurements as `trials'. Participants were briefed about all details of the study, signed a consent form, and wore laser safety goggles throughout the entire experiment. The study protocol was approved by the Duke Campus Institutional Review Board (IRB) prior to the experiments (IRB Protocol Number 2022-0093). Demographics of all subjects are shown in the supplementary material.

\paragraph{PDCS measurement at the forearm during blood flow suppression}
Subjects placed their right forearm on a table, and the optical setup consisting of three fibers was placed on the arm. A standard blood pressure cuff was placed on the same arm of the subject. Three independent trials were recorded and each trial consisted of an initial control measurement without occlusion, one test measurement during occlusion, and a second control measurement seconds after the release of the pressure. Each measurement interval lasted 7.5~s, with a break of about 10~s between two measurements (i.e., between control and occlusion measurement) to save the raw data to the drive. From the total number of 24 subjects, complete data from both detectors was available for 33 trials from 11 subjects. A pre-defined exclusion criterion for the residual of BFI estimation was applied to this data and resulted in the removal of nine trials, as shown in our supplementary material. This automated removal of extreme outliers followed the pre-registered analysis plan at \url{https://doi.org/10.17605/OSF.IO/TJ2WV}.

\paragraph{PDCS measurement at the prefrontal cortex during cognitive activation}
For these measurements, subjects placed their head on a chin rest. Three optical fibers were positioned at the forehead at 1~-~2~cm above the right eyebrow. A small monitor was placed at about 20~cm distance at eye level of the subject and their hands were placed on a keyboard to control the required keys for the cognitive task. The n-back task (n~=~2) was carried out on a web service provided by PsyToolkit’s experiment library at \url{https://www.psytoolkit.org/experiment-library/nback2.html}~\cite{stoet2010psytoolkit,stoet2017psytoolkit}. This task is know to cause activation in the PFC, which is responsible for short-term memory and action planning~\cite{kane2007working}. During this task, a series of 25 letters is shown to the participants, usually containing six instances, where the same letter is repeated after two letters (see Figure~\ref{Fig_2}~c). If subjects correctly indicated this instance, it was counted as a correct match. Missed instances and false alarms were also counted. Prior to the actual experiment, each subject was instructed in the procedure and the rules of this n-back test. Each subject carried out one practice test, without data acquisition to familiarize themselves with the test. For all subsequent trails with data acquisition, the percentage of correct matches, the percentage of missed matches, and the percentage of false alarms were documented, as shown in the supplementary material. Each trial consisted of an initial control measurement, followed by a measurement during the task and, finally, another control measurement (see Fig.~\ref{Fig_2}~a). The actual n-back test lasted about 40~-~50~s and the PDCS measurement was started about 15~-~20~s after the start of the test. Each PDCS measurement lasted 15~s. After the task was completed, subjects were again asked to close their eyes and a second control measurement was recorded. There was a break of about 20~-~30~s between two measurements in the same trial (i.e., between control measurement and task measurement). The same procedure was repeated three times, to obtain three independent trials for each subject. From the total number of 24 subjects, complete data from both detectors was available from 48 trials. Nine trials of the cognitive task were excluded based on the pre-defined exclusion criterion for the residual of the BFI reconstruction for a total of 39 trials from 15 subjects, as described in detail in our supplementary material.

\subsection*{Optical setup}
As displayed in Fig.~\ref{Fig_1} and in Fig.~\ref{Fig_2}~b~\&~c, we used a dual PDCS system, with one laser source ($\lambda~=~785nm$) and two individual detection setups. The laser was operated at 100~mW and connected to a large fiber patch cable (Thorlabs M107L, 1500~\textmu m, NA~=~0.50). An empty lens tube with a length of 16~mm was mounted to the laser fiber to act as a spacer between the fiber tip and the subject's skin. This ensured a sufficiently large beam diameter at the skin and an intensity well below the maximal permissible exposure (MPE) limit of 300 mW/cm². The first detection setups consisted of the commercially available 32$\times$32 SPAD camera (PF-32, PhotonForce Ltd., Scotland, UK) that was connected to a 200~\textmu m fiber (Thorlabs M44L, 200~\textmu m, NA~=~0.50), at 1.5~cm distance to the laser fiber (SDS~=~1.5~cm). The second setup was based on the novel 500$\times$500 SwissSPAD3 array~\cite{wayne2022500}, connected to a 1500~\textmu m fiber (Thorlabs M107L, 1500~\textmu m, NA~=~0.50), at a distance of 4~cm to the laser fiber (SDS~=~4~cm). In both cases, the distance between the fiber end and the sensor (z) was adjusted to match the speckle diameter (d) to the active area of the detector, according to~\cite{goodman2007_book}:
\begin{equation}
    d~=~\frac{\lambda z}{D}
\end{equation}
where $\lambda$ is the wavelength (785~nm) and D is the fiber diameter. In the case of the setup at SDS~=~1.5~cm, the diameter of the active area of each pixel in the PF-32 array is 6.9~\textmu m, resulting in a distance of z~=~1.76~mm between fiber and detector. In the case of the long source-detector separation (SDS~=~4~cm), the setup followed the procedure described by Wayne et al.~\cite{wayne2023massively}, with a distance of z~=~11.5~mm between fiber and detector, since the diameter of the active area of each pixel in the SwissSPAD3 array is 6~\textmu m. To ensure a correct setup, the speckle patterns at these distances were measured by a camera with 1.85~\textmu m pixel size (acA4024-29um, Basler AG, Ahrensburg, Germany).

\subsection*{Data processing}
\label{chapter_data_processing}

DCS measurements are based on the normalized intensity autocorrelation, given by:

\begin{equation}
    g_2(\tau)~=~  \frac{\langle I(t) * I(t + \tau) \rangle}{\langle I(t) \rangle^2}
    \label{equation_g2}
\end{equation}

where the brackets $\langle \rangle$ denote time averages, $\tau$ denotes the time delay variable of the temporal autocorrelation and I(t) is the measured number of binary photon detection events. In each case, this $g_2$ was calculated based on Equ.~\ref{equation_g2} per individual pixel of the sensor. We employed two different modes of operation to obtain the $g_2$ data: (1) streaming of full raw frame data. This data was transferred via an iPASS cable (IPASS PCIE CABLE ASSY 68P 3M, 0745460801, Molex, Wellington, USA) that was connected to a PCIE cable assembly expansion board (AVT-ONIX-PCIE-IPASS-8X-G board, Onix systems, Israel), saving full frame data to a SSD hard drive or (2) automated, real-time, on-board calculation of the $g_2$ curves on the respective FPGA unit. The FPGA firmware was developed by Wayne et al.~\cite{wayne2023massively} and outputs $g_2(\tau)$ values over a USB3.0 connection at a fixed number of 16 delay times, between $\tau_0~=~0$ (no delay) and $\tau_{max}$ at 15 times the exposure time.

The temporal resolution $\Delta \tau$ of these $g_2$ curves is defined by the total exposure time of each sensor. The 32$\times$32 PF-32 detector has a global shutter and the exposure time was held constant at 3~\textmu s in all experiments. The larger 500$\times$500 swissSPAD3 detector has a shutter operation, where the number of readout rows can be adjusted. Thus, its full-frame exposure time is defined by the number of rows that is read, multiplied by the readout time for each row of 40~ns~\cite{wayne2022500,wayne2023massively}. In the full raw data streaming mode, the number of rows could be adjusted freely between 1 and 250, for a full-frame exposure time of $t_{exposure}~=~n_{rows} \times 40$~ns. The on-board autocorrelation mode only allowed measurements of either 250 rows from one sensor half at 10.81~~\textmu s exposure (250~$\times$~40ns~=~10~\textmu s, plus additional wait time for wait states and clock cycle) or 64 rows from two sensor halves (128 rows in total) at 3.37~\textmu s exposure (64~$\times$~40~ns~=~2.56~\textmu s, plus additional wait time). For the experiments described in this work, we used two different settings, specifically for each of the two experiments, as addressed in the discussion section of this paper. In each case, $g_2$ was calculated for each individual SPAD pixel and then all curves were averaged across all pixels and across the temporal averaging window $\Delta T$ to boost the SNR.
An overview of these settings is summarized in the supplementary material. 

\paragraph{BFI estimation}
All calculated $g_2$ curves were used to estimate the blood flow index (BFI), according to the semi-infinite solution to the correlation diffusion equation, which has been described extensively in literature~\cite{boas1995scattering,boas1997spatially,cheung2001vivo,culver2003diffuse,durduran2005diffuse,Nakabayashi2023}. We used publicly available code for this model from Wu. et al.~\cite{wu2023scatterbrains}. Please refer to our supplementary material for more details. In brief, a closed-form, analytical model for the normalized electric field autocorrelation function  $g_1$ based on homogeneous tissue properties was used to model the BFI as the diffusion coefficient. This expression for $g_1(\tau)$ was then inserted into the Siegert relation~\cite{lemieux1999investigating}:
\begin{equation}
g_2(\tau)~=~1 + \beta*|g_1(\tau)|^2
\label{equation_siegert}
\end{equation}
This analytical expression for the normalized intensity autocorrelation $g_2(\tau)$ was then fitted to the measured $g_2(\tau)$ curve (see Equ.~\ref{equation_g2}) and optimized for the BFI and $\beta$ in each individual $g_2$ of a given time point $\Delta T$. In the case of the experiment to measure cerebral blood flow at large SDS, only the initial 12 delay values ($\tau$~=~3.3~-~40.4~\textmu s) were considered for the BFI estimation, as the later delay was more noisy, as discussed in the Results section. Finally, the goodness of the fitted semi-infinite diffusion correlation model was analyzed by considering its residuals. The mean residual per trial was analyzed and a threshold for the median residual of 0.03 was used to exclude outliers from further statistical analysis described below.

\paragraph{Additional blood flow metrics}
Besides the established BFI, we tested the initial $g_2$ difference between the first two decay values:
\begin{equation}
    \Delta g_{2,init}~=~g_2(\tau_1) - g_2(\tau_2)
\end{equation}
The rationale for this metric is that it carries information on changes in speckle contrast, similar to (r)BFI, but focusing solely on the two earliest delay values at highest blood flow sensitivity and SNR. Since the $g_2$ curve decayed relatively fast at this large SDS (see Fig.~\ref{Fig_3}~and~\ref{Fig_4}), it was tested for blood flow sensitivity. This initial $g_2$ difference was sampled at the same rate as the BFI (8~Hz for the cognitive activation experiment and 10~Hz for the suppression experiment). 

\paragraph{Pulsatility analysis}
In this section, we describe the algorithm for measuring pulse frequency as well as pulse markers for the calculation of the pulsatility index and resistance index. 

First a fast Fourier transform (FFT) was carried out of the entire BFI trace within the upper and lower frequency boundaries of 0.33~-~2.65~Hz or 20~-~160~bpm. Then Matlab's findpeaks function was applied on the FFT data and the peak of second highest probability (the first secondary peak) was identified as the pulse frequency. 

Then systolic peaks (SP) were identified as local maximal in the BFI trace, using Matlab's findpeaks function with a minimal distance of 0.6 times the previously determined pulse rate. Diastolic end points (DE) were then obtained as minima in the BFI in between two subsequent systolic peaks and diastolic notches (DN) were measured as local minima within a window of 0.4~s following a systolic peak. To ensure that there is the same number of peaks, notches and end points, the first or last SP was ignored, if there was an additional entry. Diastolic peaks (DP) were then identified as local maxima between subsequent diastolic notches (DN) and diastolic end points (DE). In order to prevent a faulty analysis in cases, where the actual flow rate was too fast for the sampling rate to detect the diastolic peak accurately, all points where a diastolic peak and a diastolic notch fell on the same point were removed. Finally, the pulsatility index (PI) was calculated as:
\begin{equation}
     PI = \frac{ \langle BFI \rangle_{SP} - \langle BFI \rangle_{DE}}{\langle BFI \rangle}
\end{equation}
Where $\langle BFI \rangle$ is the average BFI of the entire trial, while $\langle BFI \rangle_{SP}$ and $\langle BFI \rangle_{DE}$ denote the respective averages across all detected peaks or end points. Additional pulse metrics are displayed in the supplementary material.

\subsection*{Noise analysis}
The above mentioned detection of pulse markers also enables a noise estimation of these \textit{in vivo} data, by considering all peaks (SP) in a BFI trace as subsequent measurements under similar blood flow conditions (all end points can serve as an additional set of similar blood flow conditions respectively). We then calculated the the average noise $\overline{N}$ of each individual trial as the standard deviation (STD) of each $g_2(\tau)$ value, across n data points under similar conditions (i.e., systolic peaks or diastolic end points respectively):
\begin{equation}
    \overline{N} = \langle std(g_2(\tau))_n \rangle_\tau 
\end{equation}
where $g_2(\tau)$ is the autocorrelation function, averaged across all pixels in the detector. The noise was then averaged across all delay values, denoted by the brackets $\langle N(\tau) \rangle _\tau$. The distribution of these trial-based noise values was then plotted as boxplot, as seen in Fig.~\ref{Fig_8}. A more detailed noise analysis, shown as a function of $\tau$ and different pixel counts $M_{pixel}$, following the $\sqrt{M_{pixel}}$ reduction in noise can be found in the supplementary material.

\subsection*{Statistical analysis}
For each trial, the BFI and the initial $g_2$ slope were normalized to 100~\% of the median of the first control measurement. The data from each of the two detectors was analyzed separately so that the short source-detector separation (SDS~=~1.5~cm) could serve as an in-built control measurement for the deep PDCS configuration (SDS~=~4~cm). 

For each individual trial, the median of the time course was calculated for the first control, the test condition (suppression by cuff or activation by cognitive task), and for the second control. The group distribution of these median values in all trials was plotted and analyzed by a linear mixed effects (LME) model to derive p and t values. The LME used the normalized, averaged blood flow metric (relative BFI - rBFI) as the outcome variable while using the trial and the condition (control or experimental) as predictor variables with a random intercept and slope for each trial:
\begin{equation}
    rBFI \sim Trial+Condition+(Condition|Subject)
\end{equation}
We ran the same analysis for both sensors to compare blood flow changes that were recorded simultaneously from superficial and deeper tissue layers. Furthermore, the average difference and Cohen's difference~\cite{Bettinardi2023} of the distribution were calculated. The same procedure was also applied to compare pulsatility parameters between the two detectors. This analysis procedure follows the pre-registered study design at \url{https://doi.org/10.17605/OSF.IO/TJ2WV}.

\bibliography{sample}

\section*{Acknowledgements}
R.H. acknowledges support from a Hartwell Foundation Individual Biomedical Researcher Award. This material is based upon work supported by the Air Force Office of Scientific Research under award number FA9550-21-1-0401. Mi.Wa. acknowledges partial support by the Swiss National Science Foundation (grant 20QT21-187716 Qu3D “Quantum 3D Imaging at high speed and high resolution”) and by Meta Platforms Inc. We thank Nir Beckermus and his team at Beckermus Technologies ltd for their support during this project. Furthermore, we are grateful to Dr.-Ing. Paul Ritter and Dr. Kyle Cowdrick for the valuable discussions on blood flow dynamics, as well as to Dr. Kevin Zhou for valuable feedback. 

\section*{Author contributions statement}
Conceptualization: L.K., M.W., Mi.Wa., S.X., P.M., D.D., S.H., R.W.; Methodology and Experimental setup: L.K., M.W., Mi.Wa., S.X., K.K., M.H., R.H.; Conduction of experiments: L.K.; Data analysis: L.K., M.W.; Visualization: L.K., M.W., K.C.L., W.L., R.W.; Supervision: S.A.L., E.B., C.B., E.C., S.H., R.H.; Writing—original draft: L.K.; Writing—review \& editing: all other authors \\


\section*{Additional information}

\subsection*{Disclosures} 
M.W., R.H. and L.K. have submitted a patent application related to this work, assigned to Duke University. Edoardo Charbon holds the position of Chief Scientific Officer of Fastree3D, a company making LiDARs for the automotive market, and Claudio Bruschini and Edoardo Charbon are co-founders of Pi Imaging Technology. Neither company has been involved with the work or paper. The authors declare no other competing financial interests. All other authors declare no conflicts of interest. 

\subsection*{Data Availability Statement} 
The code for the diffusion model was based on the scatterBrains project from Wu. et al.~\cite{wu2023scatterbrains}, which is publicly available at \url{https://github.com/wumelissa/scatterBrains}. Due to the immense size of the raw data, the data is only available from the corresponding author upon reasonable request.

\end{document}